# Excitonic effects on the third-order nonlinear optical properties of solids: Theory and application

You-Zhao Lan*[1]

*Key Laboratory of the Ministry of Education for Advanced Catalysis Materials, College of Chemistry and Life Sciences, Zhejiang Normal University, Jinhua, Zhejiang, 321004, China*
*State Key Laboratory of Structural Chemistry, Fuzhou, Fujian, 350002, China*

**Abstract**

We present a many-body Bethe-Salpeter equation eigenstates based sum-over-states method to calculate the linear and nonlinear optical properties of solids. Excitonic and local field effects are included in the calculations. As applications, we calculate the one-photon absorption, third harmonic generation, degenerate four-wave mixing spectra of solid $C_{60}$ fullerene. The overall agreement between the theoretical and experimental results is very good for all three calculated spectra. By comparisons with the independent particle approximation based sum-over-states method, we show that excitonic effects mix the independent particle transition peaks to new excitonic ones. The position and intensity of spectral peaks are modified significantly. By tracing the sum-over-states progress, we determine the type of nonlinear polarization resonances for the characteristic peaks of third harmonic generation process, which may clear up a discrepancy in two experimental results.

## 1. Introduction

Excitonic effects are important for calculating the linear and nonlinear optical properties [1–8]. In the linear optical calculations, excitonic effects lead to an excellent agreement between theoretical and experimental results in terms of peak positions and intensities [5,8]. In the nonlinear optical calculations, excitonic effects significantly modify the intensity and distribution of peaks in the third-harmonic generation of carbon nanotubes and graphene nanoribbons [1]. Excitonic effects are also found to redistribute the oscillator strength in the second-harmonic polarizabilities [2,6]. Here, we will develop the sum-over-states (SOS) method based on the Bethe-Salpeter equation (BSE) eigenstates (BSE-SOS) to calculate the nonlinear optical properties of solids, which include the excitonic and local field effects naturally.

---
[1] Corresponding author: lyzhao@zjnu.cn

The SOS method based on the standard perturbation technique has been widely used to calculate the linear and nonlinear optical properties of molecular systems [9–17]. Compared to other methods, such as the finite-field method, analytic derivative method, and response theory, the SOS method has an obvious advantage that we can easily calculate the dynamic optical properties and analyze microscopic contributions to the optical properties. In principle, the SOS method can be combined with any electronic structure method which can determine the wave functions and transition moments of approximate states of a system. So the performance of the SOS method to some degree depends on the accuracy of the electronic structure method. However, the convergence issue of the SOS method limits the combination with the highly accuracy electronic structure methods. Various electronic structure methods, from semi-empirical [9,11,12,15,18] to *ab-initio* or density functional methods [19,20], have been used to improve the understanding of electronic contributions to the optical properties of molecular system. Up to now, the semi-empirical method is still preferable for large molecules [21].

For solids, similar strategies have also been developed to calculate the linear and nonlinear optical properties within the independent particle approximation (IPA) [22–27]. Within IPA, the expressions for calculating optical properties sum over the energy bands of solids, we call this method the sum-over-bands (SOB) method which has been widely used for semiconductors [28–30]. The SOB method is similar to the summation over orbitals (SOO) method [31] in molecular systems which is more approximate than the SOS method and rarely used. According to researches on the one-photon absorption (OPA) of solids [8], calculations based on the summation over excitonic states (here also called SOS corresponding to that in molecular calculations) lead to an excellent agreement between theoretical and experimental results. Similar computational strategies also lead to an improved understanding of the two-photon absorption (TPA) of solids [5]. The excitonic states obtained by solving the many-body BSE are apparently more accurate than the single particle states based on the IPA to describe nonlinear optical response of solids, so the better theoretical results are expectable.

In this paper, the BSE-SOS method is first to solve the self-consistent Kohn–Sham equations with the generalized gradient approximation (GGA) of the Perdew–Burke–Ernzerhof (PBE) functional [32] combined with the all-electron full-potential linearised augmented-plane wave (FLAPW) method [33], and then to solve the BSE to obtain the excitonic states of system, finally based on the BSE eigenstates, we use the SOS expressions to calculate the optical susceptibilities. Note that a similar BSE eigenstates based SOS calculation has been performed by Leitsmann *et al.* [6] for calculating the second harmonic generation of semiconductors.

In their method, the velocity-gauge perturbation was used and the final SOS expressions are based on the momentum matrix elements and excitonic transition energy. Use of velocity-gauge leads to unphysical divergences at zero frequency in the SOS expressions, similar to the IPA based SOB method. This problem can be solved by using the length-gauge perturbation which requires the estimation of position matrix element. Although the dipole or position operator of a solid is ill defined in periodic systems [34], a length-gauge perturbation compared to a velocity-gauge one leads to less abstract and better understandings of nonlinear optical physical issues, and more importantly to the avoidance of unphysical divergences at zero frequency [22,25,35]. Since the initial BSE states are linearly expanded by the single particle states obtained within IPA, the position matrix elements of the BSE states depend on those of the single particle states, similar to the calculation method of the BSE states based momentum matrix elements used by Leitsmann *et al.* [6] and Chang *et al.* [2]

We implemented our strategy in the revised ELK program [36]. As applications, we calculated the linear and nonlinear optical properties of solid $C_{60}$ fullerene. Overall, the calculated OPA and the third harmonic generation (THG) spectra agree well with the experimental ones. The calculated degenerate four-wave mixing (DFWM) spectrum is in excellent agreement with the available experimental results in terms of line shape and peak positions. Meanwhile, the physical origin of the optical responses of $C_{60}$ has been discussed by tracing the SOS process.

In Section 2, we describe theoretical framework of the BSE-SOS method. In Section 3, we show the computational details, after which we discuss the calculated OPA, THG, and DFWM spectra of $C_{60}$ solid fullerene and make comparisons between the theoretical and experimental results. Finally, conclusions are given in Section 4.

## 2. Theory

We first outline how to obtain the optical susceptibility based on the perturbation solution of the time-dependent Schrödinger equation, a detailed description can be found elsewhere [37,38]. The evolution of states of a system can be described in terms of the following time-dependent Schrödinger equation.

$$i\hbar \frac{\partial \psi}{\partial t} = H\psi \qquad (1)$$

Eq.1 cannot be solved exactly for a system exposed to an electromagnetic field. Using the standard perturbation method, we split the Hamiltonian operator (Eq. 1) as

$$H = H_0 + V(t) \qquad (2)$$

, where $H_0$ is the unperturbed Hamiltonian operator, $V(t)$ describes the interaction caused by perturbation. For the optical susceptibility, we introduce the electron-radiation interaction within the electric dipole approximation, that is,

$$V(t) = -\mu E(t) \qquad (3)$$

, where $\mu$ is the electric dipole moment ($\mu = -|e|r$) and $E(t)$ is the electric field. For a system exposed to the electron-radiation, the optical polarization ($P$) can be given by (for simplicity, in the scalar form)

$$P(\omega,t) = \varepsilon_0 \chi^{(1)}(\omega) E(\omega,t) + \varepsilon_0 \chi^{(2)}(\omega) E^2(\omega,t) + \varepsilon_0 \chi^{(3)}(\omega) E^3(\omega,t) + \cdots \qquad (4)$$

, where $\chi^{(1)}$, $\chi^{(2)}$, and $\chi^{(3)}$ are the first-, second-, and third-order optical susceptibilities, and $N$ is the number density of units. In terms of perturbation solution of Eq. 1, the optical susceptibilities can be calculated by [37,38]

$$\chi_{ij}^{(1)}(\omega_p;\omega_1) = \frac{N}{\varepsilon_0 \hbar} \sum_m \left( \frac{\mu_{gm}^i \mu_{mg}^j}{\omega_{mg} - \omega_1} + \frac{\mu_{gm}^j \mu_{mg}^i}{\omega_{mg} + \omega_1} \right) \qquad (5a)$$

$$\chi_{ijkh}^{(3)}(\omega_p;\omega_1,\omega_2,\omega_3) = \frac{N}{\varepsilon_0 \hbar^3} I_{\omega_1 \omega_2 \omega_3}^{j,k,h} \sum_{mnv} \left( \begin{array}{c} \frac{\mu_{gv}^k \mu_{vn}^j \mu_{nm}^i \mu_{mg}^h}{(\omega_{vg} - \omega_1 - \omega_2 - \omega_3)(\omega_{ng} - \omega_2 - \omega_3)(\omega_{mg} - \omega_p)} + \frac{\mu_{gv}^j \mu_{vn}^k \mu_{nm}^i \mu_{mg}^h}{(\omega_{vg} + \omega_1)(\omega_{ng} - \omega_2 - \omega_3)(\omega_{mg} - \omega_3)} \\ + \frac{\mu_{gv}^j \mu_{vn}^i \mu_{nm}^k \mu_{mg}^h}{(\omega_{vg} + \omega_1)(\omega_{ng} + \omega_1 + \omega_2)(\omega_{mg} - \omega_3)} + \frac{\mu_{gv}^j \mu_{vn}^i \mu_{nm}^h \mu_{mg}^k}{(\omega_{vg} + \omega_1)(\omega_{ng} + \omega_1 + \omega_2)(\omega_{mg} + \omega_p)} \end{array} \right) \qquad (5b)$$

The second-order optical susceptibility $\chi^{(2)}$ that has a similar expression is not shown here because we will not use it in present work. In Eq. 5, the indices of $i$, $j$, $k$, $h$ are the Cartesian directions, $\omega_1$, $\omega_2$ and $\omega_3$ are the applied field frequencies, $\omega_p$ is the polarization frequency ($\omega_p = \omega_1$ and $\omega_p = \omega_1 + \omega_2 + \omega_3$ for Eq. 5a and Eq. 5b, respectively), $\mu_{gm}$ and $\mu_{nm}$ are the transition dipole moments between the ground ($g$) and excited ($m$) states and between the excited states ($m$ and $n$), respectively, and $\omega_{mg}$ is the transition energy of excited state of system. To consider the lifetime broadening or population decay rates ($\eta$) of the excited state, we have implicitly defined $\omega_{mg}$ as $\omega_{mg} + i\eta/2$. When $m = n$, $\mu_{nm}$ indicates the difference in the dipole moment between the ground and excited states. The intrinsic permutation operator ($I$) indicates the permutations between $\omega_1$, $\omega_2$, and $\omega_3$, and the corresponding Cartesian indices ($j$, $k$, and $h$) are to be permuted simultaneously.

In principle, we can calculate the optical susceptibilities/polarizabilities based on Eq. 5, provided that we obtain the transition dipole moments and excitation energies of excited states of a system. For molecules, Eq. 5 has been widely used to calculate the linear and nonlinear optical polarizabilities [9,12,16,39]. The standard electronic structure methods (Hartree-Fock, configuration interaction, multi-configuration self consistent field, etc.) were used to estimate the transition moments and excitation energies of approximate states of a system. For solids, within the IPA the transition matrix elements of the position operator between different energy bands ($s$, $t$, etc.) can be calculated by using the finite difference approach or relating the position operator to the velocity operator [8]. In our present work, we prefer to the latter, that is, we use the following relation,

$$r_{st} = \langle s|r|t\rangle = \frac{\langle s|[H_0,r]|t\rangle}{\hbar\omega_{st}} = \frac{\langle s|\upsilon|t\rangle}{i\omega_{st}} = \frac{\langle s|p|t\rangle}{im\omega_{st}} \quad (6)$$

, where $r_{st}$ can be set to zero if $\omega_{st} = 0$ (see Ref.28 of Ref.[40]). Note that the third equality of Eq.6 is accurate only when the potential commutes with position operator (i.e. $[V(r),r] = 0$). In the case of nonlocal pseudopotentials, the third equality of Eq.6 strictly holds no longer. Here, this equality strictly holds because we will use the all-electron FLAPW method.

We use the effective two-particle Hamiltonian ($H_0^{2p}$ in Eq.2) based on the BSE, which makes the electron-hole interaction and the local field effect be included in the calculation. The $H_0^{2p}$ can be defined as [8,41]

$$H_0^{2p} = H_0^{1p} + H_0^{eh} \quad (7)$$

, where $H_0^{1p}$ is the independent-particle contribution simply given by the band-structure energy differences $E_{ck}-E_{vk}$, and $H_0^{eh}$ is the electron-hole interaction term which includes the direct attraction interaction term ($H_0^{eh,d}$) and the exchange term ($H_0^{eh,x}$). While the $H_0^{1p}$ describes the independent-particle excitation, the $H_0^{eh}$ means the coupling between different independent-particle transitions (*valence band* ($v$)→*conduction band* ($c$)). In this case, the eigenstates of $H_0^{2p}$ indicate the electron-hole excited states $|m\rangle$ (*i.e.*, $H_0^{2p}|m\rangle = \hbar\omega_m|m\rangle$). The $|m\rangle$ is given by the linear combination of independent particle excitations $|vck\rangle$ (*i.e.*, $|vk\rangle$ to $|ck\rangle$) as

$$|m\rangle = \sum_{c,v,k} A_{c,v,k}^m |vck\rangle \quad (8)$$

The $\langle g|r|m\rangle$ and $\langle m|r|n\rangle$ in Eq. 5 can be computed in terms of the general rule for the matrix element of one-particle operator between Slater determinants [42,43], that is,

$$\langle g|r|m\rangle = \sum_{c,v,k} A_{c,v,k}^m r_{cv,k}^{IPA} \quad (9)$$

$$\langle m|r|n\rangle = \sum_{c_1,c_2,v_1,v_2,k} A_{c_1,v_1,k}^{m*} A_{c_2,v_2,k}^{n} \left( r_{c_1c_2,k}^{IPA}\delta_{v_1v_2} + r_{v_1v_2,k}^{IPA}\delta_{c_1c_2} \right) \quad (10)$$

## 3. Applications

3.1 *Computational detailed.*

We constructed the experimental face-centered cubic (fcc) crystal structure [44] of solid $C_{60}$ fullerene. The structure was optimized using the density functional theory within the GGA-PBE combined with the pseudopotential plane wave method, as implemented in the CASTEP code [45] of Material Studio 4.0. A k-point mesh of 4×4×4, force threshold of 0.01 eV/Å, and stress threshold of 0.02 GPa were used for the optimizations. The optimized lattice parameter (*a*) is 14.26 Å, which agrees with the experimental lattice parameter (14.17 ± 0.01 Å) and with the theoretical result [46] based on the calculation within the GGA-PBE combined with the FLAPW method [33]. The optimized structure along the [110] direction is shown as an inset in Fig. 1.

The optimized structure was used for the band structure calculation. We performed the band structure calculation by using the GGA-PBE combined with the all-electron FLAPW method, as implemented in ELK code [36]. A *k*-point mesh of 4×4×4 was used for the band structure calculation.

For the optical properties, the energy bands within the IPA were obtained by solving the self-consistent Kohn–Sham equations with the GGA-PBE functional, and the excitonic states were obtained by solving the BSE with a basis linearly expanded by the IPA states (Eq.8). The corresponding position matrix elements were also calculated by the revised ELK which reads $r^{IPA}$ to calculate the BSE states based position matrix elements in terms of Eqs. 9 and 10. As the GGA-PBE calculation generally underestimates the band gap of solid, the scissor correction was used and the corresponding scissor value was based on the difference between the theoretical and experimental electronic energy gaps. Our theoretical energy gap is 1.14 eV at X point (Fig. 1), which agrees with the previous theoretical results [46] at the same level of theory. The adopted experimental gap [47] is 2.14 eV close to the energy gap of 2.15 eV based on the GW calculation [48].

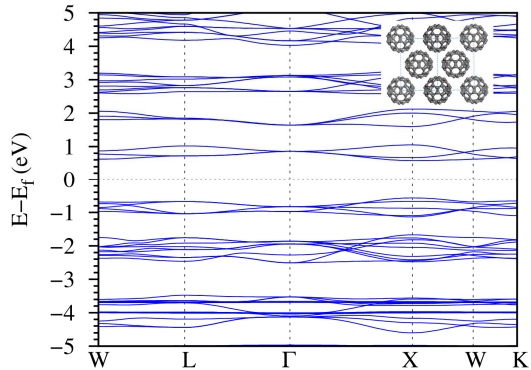

**Fig. 1.** *Band structure of $C_{60}$ with the face-centered cubic crystal structure (inset along [110] direction). The Fermi energy is set to zero.*

To compare our theoretical results with the experimental ones [49], we calculated the third-order polarizability for the THG and DFWM processes, that is, $\chi^{THG}(\omega) = \chi^{(3)}_{iiii}(-3\omega;\omega,\omega,\omega)$ and $\chi^{DFWM}(\omega) = \chi^{(3)}_{iiii}(-\omega;\omega,-\omega,\omega)$ ($i = x, y, z$) were calculated. The fcc-$C_{60}$ structure with a $T_h$ symmetry has an identity [37] of $\chi^{(3)}_{xxxx} = \chi^{(3)}_{yyyy} = \chi^{(3)}_{zzzz}$. Since the linear absorption has an significant impact on the nonlinear optical process [49], we also calculated the OPA spectrum using the same level of theory and compared it with the experimental one. In the BSE calculation, we performed the convergence tests on the k-grid, the number of valence and conduction states included in the BSE, and the number of excitonic states included in the SOS calculations. As an example, the calculated THG results are shown in Fig. 2, where the experimental ones are included for comparisons. Since the BSE calculation is very expensive, for the fcc-$C_{60}$ with 300 empty states included the response function, we can only run the k-grid of 4×4×4 calculation at most owing to the limit of computational resources. As shown in Fig. 2a, with a fixed v10c15 pair, the OPA spectrum based on the k-grid of 4×4×4 has been in good agreement with the experimental one in terms of peak position and line shape. The k-grids of 1×1×1 and 2×2×2 are not dense enough to produce the OPA spectra. For the THG spectra (Fig. 2b), with the peak position shifted by 0.1 eV and the values divided by 5, the k-grid of 4×4×4 almost reproduces the experimental THG spectrum. Similarly, the k-grids of 1×1×1 and 2×2×2 are obviously not dense enough for the THG spectra. In Figs. 2c and 2d, we examined the dependence of OPA and THG spectra on the number of valence (v) and conduction (c) states included in the BSE calculation. It is shown that both the OPA and THG spectra converge in the v10c15 pair.

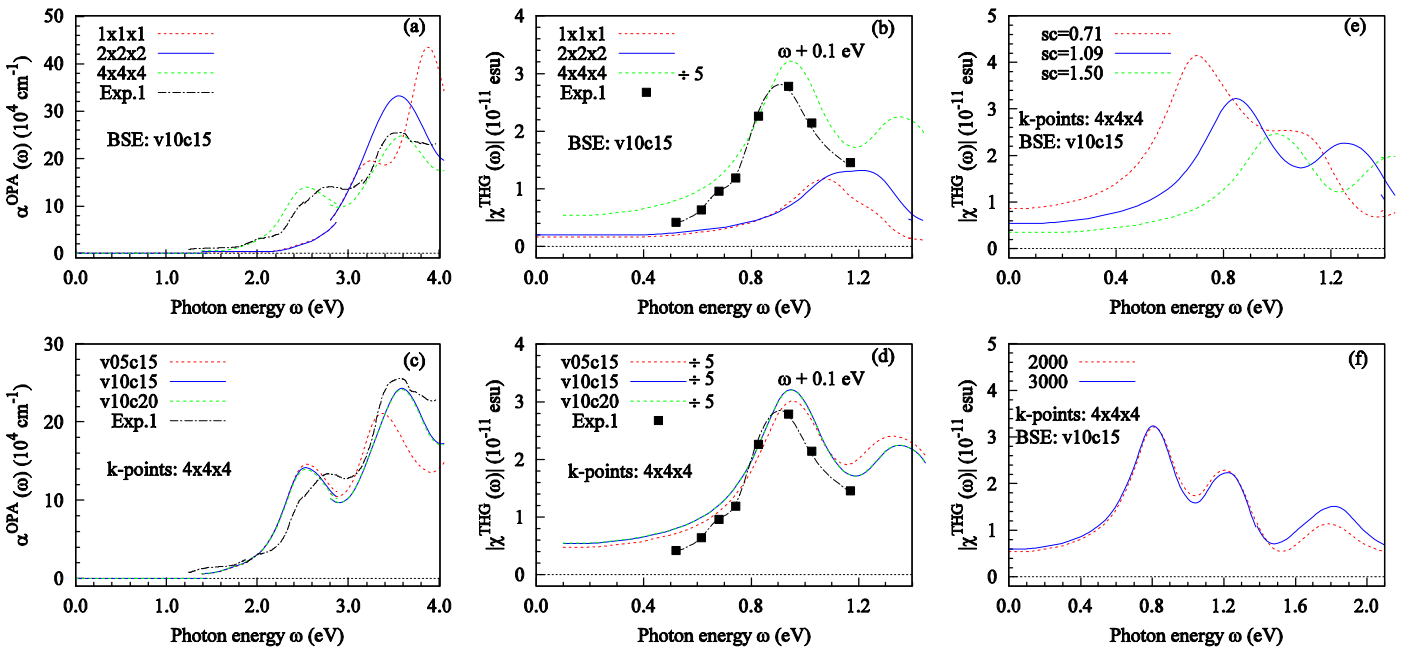

**Fig. 2.** *Dependences of (a and c) OPA and (b and d) THG on the k-grid and the number of valence (v) and conduction (c) states included in the BSE. The legend of v05c15 indicates the 5 highest valence states and 15 lowest conduction states are included in the BSE calculation. To fit to experimental THG spectra, (b and d) the theoretical THG spectra based on the 4×4×4 k-grid are shifted right by 0.1 eV and the $\chi^{THG}(\omega)$ values are divided by a factor of 5. Dependence of THG on (e) the scissor value and (f) the number of excited states included in the SOS calculation.*

As for peak shift, it should be noted that the scissor correction in the BSE calculation mainly leads to the rigid shift of peak positions and hardly change the peak intensities for OPA [50,51]. This is not the case for THG, we show in Fig.2e the scissor value dependence of $|\chi^{THG}(\omega)|$ based on three scissor values (i.e., 0.71, 1.09, and 1.50 eV based on the reported gap [46,47,52,53]). We can see that the shifted band gap not only leads to the rigid shift of peak positions (similar to OPA) but also changes the peak intensities. Finally, Fig. 2f shows the dependence of $|\chi^{THG}(\omega)|$ on the number of excited states included in the SOS calculation. For the combination of the 4×4×4 k-grid and v10c15 pair, we will obtain 9600 (4×4×4×10×15) excited states by solving the BSE. With the limit of the computational resources, we cannot include in the SOS calculation all the excited states which lead to (9600×9600×9600)-time summations in Eq. 5b. As shown in Fig. 2f, the summation on 2000 excited states produces the converged $|\chi^{THG}(\omega)|$ with the applied photon energy less than 1.5 eV which is much larger than the laser wave length (e.g. 1064 nm / 1.16 eV) used in experiments [49,54,55]. In the following sections, we will further discuss the OPA, THG, and DFWM spectra based on the 4×4×4 k-grid, v10c15 pair, and the scissor correction of 1.09 eV.

### 3.2 OPA and THG

Figure 3 shows the theoretical and experimental spectra [49,54]. The THG spectra are plotted at the third harmonic energy (3ω) for convenience of comparison with the OPA spectra. The theoretical spectra based on the IPA are calculated by using the expressions reported elsewhere. [22,28] Firstly, below 4.0 eV, there are two characteristic peaks in the BSE-OPA spectra whose positions are marked explicitly. The second peak (3.55 eV) of the theoretical spectrum is very close to those of two experimental ones (3.57 eV in Fig. 3c and 3.63 eV in Fig. 3d) whereas a deviation is observed for the first peak between theoretical (2.45 eV) and two experimental spectra (2.80 in Fig. 3c and 2.72 eV in Fig. 3d). For the IPA-OPA spectra, if the first peak at 2.80 eV is matched with the experimental ones (2.80 in Fig. 3c and 2.72 eV in Fig. 3d), an obvious deviation is observed for the second peak at 4.05 eV compared to the experimental peaks (3.57 eV in Fig. 3c and 3.63 eV in Fig. 3d). These deviations cannot be well explained here because an agreement appears in the second peak and different relative positions are shown in both peaks in two experimental spectra, that is, Exp.1 is higher than Exp.2 for the first peak (2.80 > 2.73) but lower for the second peak (3.57 < 3.63), and thus a simply rigid shift by scissor correction cannot improve the agreement between theoretical and experimental spectra.

Compared to the IPA-OPA spectrum, the BSE-OPA one has an obvious red-shift (0.35 eV for the first peak and 0.50 eV for the second peak) which indicates the role of excitonic effect in the OPA. The first peak at 2.80 eV in the IPA-OPA corresponds to the transition between the highest valence and lowest conduction bands at $\Gamma$ (Fig. 1, note that the scissor correction is 1.09 eV). As will be shown below, this transition also dominates the first excitonic absorption peak at 2.45 eV, which indicates a binding energy of 0.35 eV in the vertical transition. In spite of deviations in peak positions, the theoretical OPA spectrum, similar to two experimental ones is well matched with the corresponding THG spectra in terms of peak positions. As pointed out by Meth *et al.* [49], the optical transition at 2.8 eV (the first peak in Fig. 3c) dominates the first nonlinear peak while higher energy transitions contribute minimally. This behavior is in full agreement observed in our theoretical result (Fig. 3a) and the other experimental result (Fig. 3d). Moreover, a similar behavior is observed at the second peak (Fig. 3a and Fig. 3d). Thus, the states involved in the transitions near the characteristic OPA peak should dominate the dispersion of the nonlinearity.

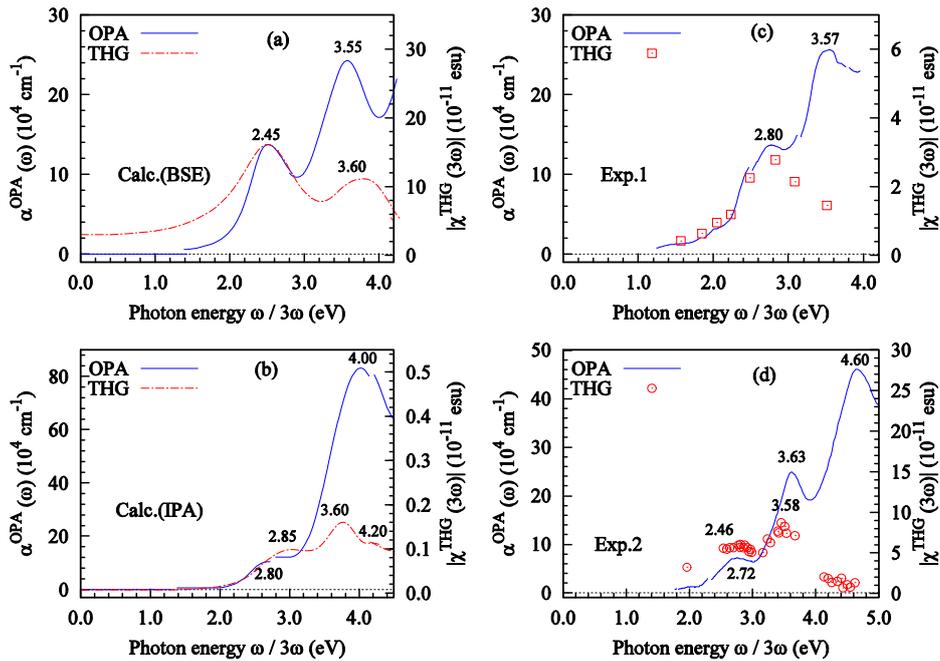

**Fig. 3.** *(a) Theoretical OPA and THG spectra based on the k-grid of 4×4×4 and the v10c15 pair (b) Theoretical OPA and THG spectra based on the IPA (pure interband transitions considered* [22,28]*). (c and d) Two experimental spectra are shown* [49,54] *for comparison. The THG spectra are plotted at the third harmonic energy (3ω) to facilitate comparison with the OPA spectrum. The photon energy of 3ω at peaks is marked explicitly.*

Secondly, the theoretical THG nonlinearity around the first peak almost tracks the linear absorption profile, and then decreases as the linear absorption increases, similar to the experimental results (Figs. 3c and 3d). The theoretical THG spectrum clearly shows the second peak at about 3.6 eV which is shown in the Exp.2 but not given in the Exp.1. Note that the Exp.1 [49] was performed on the fundamental energy range of 0.52 – 1.18 eV where the (1.18 × 3) eV just locates around 3.5 eV linear absorption peak. However, at about 3.5 eV (Fig. 3c), we cannot observe the coming of resonant peak in the THG spectrum. A possible reason is that the enhanced linear absorption would eliminate the generated third harmonic signal in experiments. The second peak was clearly shown in the Exp.2 [54] which measured more frequencies around 3.5 eV than the Exp.1. Our first theoretical peak (2.45 eV) agrees well with the peaks (around 2.46 eV) in Exp.2. Overall, the theoretical BSE-THG spectrum is in better agreement with the Exp.2 than the Exp.1 in terms of line shape and peak positions.

Thirdly, in the IPA-THG spectrum (Fig. 3b), we also observe the peaks near the OPA peaks. Since the 3ω (= 2.85 eV) is close to the OPA of 2.80 eV, this peak is mainly due to the three-photon polarization resonance. The other two peaks at 3.60 (ω = 1.20) and 4.20 eV (ω = 1.30 eV), around the OPA peak of 4.00 eV, possibly arise from both two-photon and three-photon polarization resonances because 2ω is close to the first OPA peak of 2.80 eV. As will be shown below, the peak at 3.60 eV in the BSE-THG spectrum is mainly due to the two-photon polarization resonance. In the BSE-THG spectrum, only one peak observed at around 3.60 eV is possibly due to the excitonic effect of the mixture of independent particle transitions (Eq. 8). Furthermore, excitonic effects lead to an obvious change in the magnitude of intensity, that is, the THG intensities are increased by about one order of magnitude from IPA to BSE, similar to excitonic effects on the second-harmonic polarizability (an increase in values of 20%) of semiconductors [2]. Thus, excitonic effects modify both the intensity and the distribution of peaks, also in agreement with the Attaccalite *et al.* [1] who used the real-time BSE implementation to study the nonlinear optical properties of carbon nanotube.

Finally, we observe obvious differences in the magnitude of $|\chi^{THG}(3\omega)|$ between three results. For example,

at the first peak, the theoretical |χ$^{THG}$(3ω)| is 1.6 × 10$^{-10}$ esu which is about five and three times as large as the Exp.1 (2.7 ± 0.4 × 10$^{-11}$ esu) and Exp.2 (5.5 × 10$^{-11}$ esu), respectively. Note that a direct comparison between the theoretical and experimental results is difficult because the measured results strongly depends the experimental conditions [9] (also see section 3.3). We therefore focus on the characteristic of peak position. The first OPA and THG peaks almost overlap, and do the second OPA and peaks (Figs. 3a and 3d). For the first THG peak, the Exp.1 [49] assigned it to a three-photon polarization resonance but the Exp.2 [54] assigned it to a two-photon polarization resonance. For the second peak shown in Exp.2 [54], they assigned it to a three-photon polarization resonance. So, there is a discrepancy in two experimental assignments. In our theoretical THG spectrum (Fig. 3a), the first peak at 2.45 eV can be safely assigned to a three-photon polarization resonance consistent with the Exp.1 while we assign the second peak to a two-photon polarization resonance.

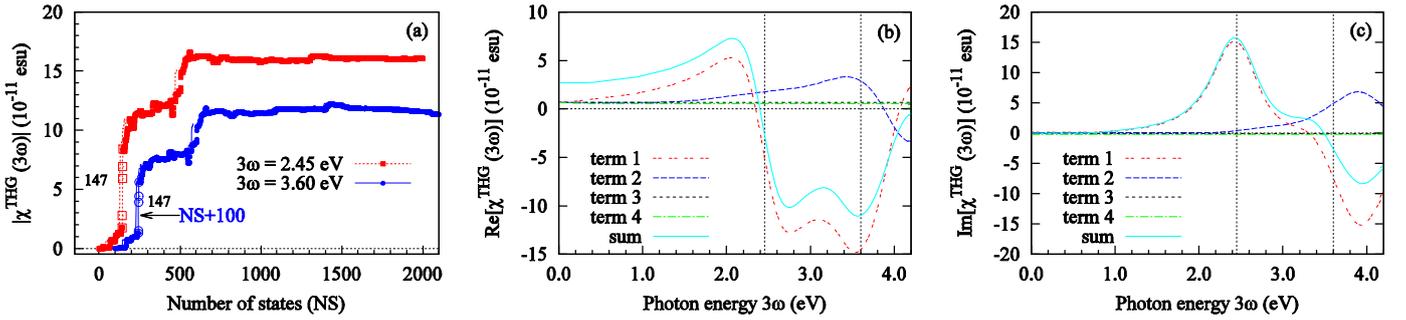

**Fig. 4.** (a) *|χ$^{THG}$(3ω)| as a function of the number of states included in the SOS calculation for 3ω = 2.45 and 3.60 eV. To avoid the overlap, the line of 3ω = 3.60 is shifted right by 100. The hollow square and circle mark the excited states with a significant contribution to the χ$^{THG}$(3ω). Decomposition of the (b) real (Re) and (c) imaginary (Im) parts of χ$^{THG}$(3ω) based on the four terms in Eq. 5b. The 'sum' indicates the summation of four terms. The two vertical dashed lines indicate the positions of two characteristic peaks (i.e. 3ω = 2.45 and 3.60 in Fig. 3a).*

To explain our assignments, we traced the SOS process. In Fig. 4a, we plot the |χ$^{THG}$(3ω)| as a function of the number of states included in the SOS calculation for 3ω = 2.45 and 3.60 eV. For 3ω = 2.45 eV, we observe that the 147$^{th}$ state has a significant contribution to the summation. The excitation energy of the 147$^{th}$ state is 2.38 eV very close to 2.45 eV, which indicates a three-photon polarization resonance. For 3ω = 3.60 eV, we also observe the 147$^{th}$ state with a significant contribution. In this case, the 2ω = 2.40 eV is very close

to the excitation energy of the 147$^{th}$ state, which indicates a two-photon polarization resonance. As shown in the denominators of Eq.5b, the two-photon polarization resonance can occur in the first and second terms while the three-photon polarization resonance only exists in the first term. Furthermore, we show in Figs. 4b and 4c the decomposition of the real and imaginary parts of $\chi^{THG}(3\omega)$ based on the four terms of Eq. 5b. As shown in Fig. 4, terms 1 and 2 have a major contribution to $\chi^{THG}(3\omega)$ with $3\omega$ larger than 1.0 eV. The term 1, which possibly leads to one-, two-, and three-photon polarization resonances, almost tracks the profile of the "sum" for both the real and imaginary parts. At $3\omega = 2.45$ eV, the imaginary part dominates the peak values of $\chi^{THG}(3\omega)$ while the real part mainly contributes to the broadening of $\chi^{THG}(3\omega)$. As the excitation energy (2.38 eV) of the 147$^{th}$ state is close to $3\omega$, this resonance is attributed to the three-photon polarization resonance. At $3\omega = 3.60$ eV, both terms 1 and 2 have a significant contribution to $\chi^{THG}(3\omega)$ by the real and imaginary parts; moreover, term 1 still has a major contribution. As shown in Eq. 5b, both term1 and term 2 can lead to one- and two-photon polarization resonances. Again, the excitation energy (2.38 eV) of the 147$^{th}$ state is close to $2\omega$, so the second peak should be attributed to the two-photon polarization resonance. As shown in Fig. 4, the 145$^{th}$, 146$^{th}$, 148$^{th}$, and 149$^{th}$, which are almost degenerate with the 147$^{th}$, have a significant contribution to $\chi^{THG}(3\omega)$. For these excited states, we list in table 1 the $\sum_{k}|A^{m}_{vck}|^2$ (summation on the k-points of the first Brillouin zone) for the major $v \rightarrow c$ transition pairs used in Eq.8. We can see that the transitions between the bands around the highest valence band and the lowest conduction band dominate in these five excited states, similar to the results based on the $C_{60}$ molecule [56].

**Table 1.** *The $\sum_{k}|A^{m}_{vck}|^2$ of the major $v \rightarrow c$ transition pairs in Eq. 8 for the selected states (see Fig. 4a) with a significant contribution to $\chi^{THG}(3\omega)$. For clarity, only the values larger than 0.1 are listed.*

|  | Ordinal of excited states | | | | |
|---|---|---|---|---|---|
| $v \rightarrow c$ | 145 | 146 | 147 | 148 | 149 |
| 118→121 [a] | 0.2509 | 0.1895 | 0.1909 | 0.1898 | 0.3738 |
| 119→121 | 0.3308 | — | 0.2686 | — | 0.2090 |
| 119→122 | 0.1257 | 0.2693 | 0.0878 | 0.2695 | 0.2111 |
| 120 [b] →121 | 0.1020 | 0.1584 | 0.1580 | 0.1584 | — |
| 120→122 | — | 0.2002 | 0.1962 | 0.1959 | 0.5446 |

[a] Ordinal of the lowest conduction (*c*) band

[b] Ordinal of the highest valence ($v$) band

3.3 *DFWM*

Besides THG, the third-order susceptibilities, $\chi^{(3)}(-\omega;\omega,-\omega,\omega)$ corresponding to the DFWM process [$\chi^{DFWM}(\omega)$], had also been measured [57–60]. Using Eq. 5b, we can easily calculate the dispersion of $\chi^{DFWM}(\omega)$. The results are shown in Fig. 5. The available experimental measurements are included for comparison. As shown in Fig. 5, our theoretical spectra are in excellent agreement with two experimental ones (Exp.3 [61] and Exp.4 [62]) in terms of line shape and peak positions. We clearly observe two nonlinear polarization resonant peaks at about 1.30 and 1.76 eV in our theoretical DFWM spectrum. The first peak (1.30 eV) agrees well with both Exp.3 (1.33 eV) and Exp.4 (1.29 eV) while the second peak (1.78 eV) only was measured by Exp.4 (1.75 eV). Note that two different factors are used to magnify the experimental values to track our theoretical results. As pointed out by Banfi *et al.* (Exp.4), their results are four times larger than Exp.3 owing to the different techniques. The first peak is safely assigned by Banfi *et al.* to the two-photon resonance due to the strong two-photon absorption. By tracing the SOS process, we can theoretically identify this peak as two-photon resonant peak due to the excited state with the excitation energy of 2.61 eV ($\approx 1.30 \times 2$). For the second theoretical peak (1.76 eV), we assign it to a two-photon resonance due to the excited state with the excitation energy of 3.50 eV, in agreement with a judgment that two-photon states at higher energy contribute to the growth of TPA observed at 3.5 eV [62].

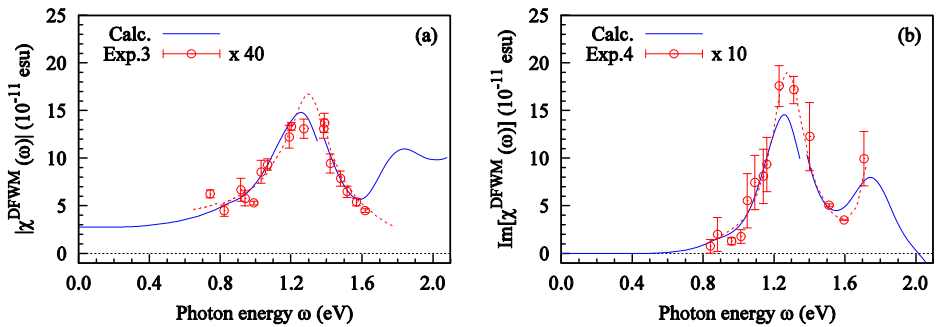

**Fig. 5**. *(a) Theoretical and experimental $|\chi^{DFWM}(\omega)|$. The experimental values (Exp.3) were based on a 10 μm amorphous $C_{60}$ film vacuum-deposited on a 1 mm $CaF_2$ substrate* [61]. *The experimental values are magnified by a factor of 40. (b) Theoretical and experimental Im[$\chi^{DFWM}(\omega)$]. The experimental values (Exp.4) were based on a polycrystalline $C_{60}$ film on a glass substrate* [62]. *The experimental values are magnified by a factor of 10. The dotted line is a Lorentzian fit based on the experimental values.*

With reference to the TPA spectrum, we notice a difficulty in identification of the OPA peak in theoretical and experimental spectra. For OPA, Exp.4 has shown an absorption peak at 2.73 eV with the half width of half maximum (HWHM) of 0.17 eV, which is consistent with the first OPA peak (~ 2.72 eV) by Exp.2 (Fig. 3d) and close to the first OPA peak (~ 2.8 eV) by Exp.1 (Fig. 3c); however, the second OPA peak (~ 3.5 eV) was obtained by Exp.1, Exp2, and our present calculation (Fig. 3) but not clearly by Exp.4. Furthermore, the first theoretical OPA peak (Fig. 3a) locates at ~ 2.45 eV which leads to a difference in the relative positions of OPA and TPA peaks. In Exp.4, the first experimental one-photon state (2.73 eV) is *higher* by 0.15 eV than the first experimental two-photon state (1.29 × 2 = 2.58 eV). However, our theoretical one-photon state at ~ 2.45 eV with a HWHM of 0.22 eV (Fig. 3a) is *lower* by 0.15 eV than the theoretical two-photon state (1.30 × 2 = 2.60 eV). Overall, agreements between theoretical and experimental spectra are obtained for two TPA peaks but a difference is shown for the first OPA peak. Moreover, we have not observed more OPA peaks around the first two-photon state. Thus, more efforts are needed to identify the OPA property around 2.60 eV.

Finally, we briefly discuss differences in magnitudes of $\chi^{THG}(\omega)$ and $\chi^{DFWM}(\omega)$ between experimental and our theoretical results. There are many discussions about differences between previous experimental and theoretical results [49,54–56,63–68]. Differences in theoretical methods and experimental techniques make comparisons difficult. In experiments, many factors, such as the type of samples, the laser pulses, and the reference standard, possibly lead to very different magnitudes of susceptibility [55,63,64], and in calculations, semi-empirical and highly accurate *ab-inito* methods will also produce different results [9]. In table 2 we list our calculated results and previous theoretical and experimental ones at a few selected wavelengths for the THG and DFWM processes. From table 2, we can observe several features: (1) Different experimental conditions, including the type of samples (film and solution) and laser pulses, lead to a large difference in values (see DFWM at 1064 nm). (2) The thickness of film affects the magnitude of $|\chi^{(3)}(\omega)|$ (see THG at 1064 nm), similar to the size-dependence of intensity of the second harmonic generation of inorganic crystals [69]. (3) Our static $|\chi^{(3)}(0)|$ value is apparently higher than previous calculations (see ∞ nm). Note that a comparison made in static values is more difficult because of no available directly measured value. The measurements cannot produce the zero-frequency $|\chi^{(3)}(0)|$ directly and must be extrapolated to zero frequency. Another main reason is that our present strategy is very different from previous calculations which mostly calculate the microscopic polarizability ($\gamma$) of gaseous $C_{60}$ molecule and then deduce the macroscopic susceptibility $\chi^{(3)}$ by using the local field factor (see caption of table 2). (4) We obtain a better agreement between calculations and experiments for THG than DFWM. A possible reason is that the DFWM process is related with the two-photon absorption which possibly leads to a large resonant measured value. Thus, although we obtain some agreements on the magnitude of $|\chi^{(3)}(\omega)|$ between theoretical and experimental results, the discrepancies still exist.

**Table 2.** *Theoretical and experimental $|\chi^{(3)}(\omega)|$ ($10^{-11}$ esu) values at a few selected wavelengths for the THG and DFWM processes. When not reported in the original work, $\chi^{(3)}$ was calculated from $\gamma$ by $\chi^{(3)} = N\gamma((\varepsilon+2)/3)^4$ based on the $N = 1.38\times10^{21}$ cm$^{-3}$ and $\varepsilon = 4$.*

| $\omega$ (eV) | THG | DFWM |
|---|---|---|
| 0 ($\infty$ nm) | 2.8, 0.088 [i], 0.075 [j], 0.011 [l], 0.035 [m] | |
| 0.65 (1907 nm) | 8.9, 3.2 [a], 0.4 [b] | 3.9 |
| 0.82 (1503 nm) | 16.1, 5.6 [a], 2.3 [b], 3.0 [g] | 5.0, 0.11 ± 0.02 [h] |
| 0.95 (1323 nm) | 10.9, 6.1 [a], 2.7 ± 0.4 [b] | 6.2, 0.14 ± 0.02 [h] |
| 1.16 (1064 nm) | 11.1, 8.2 [a], 1.5 ± 0.2 [b], 20 [d], 10 ± 8 [g] | 12.0, 0.7 [c], 330 [e], 6000 ± 4000 [f], 0.31 ± 0.03 [h] 1.6 [k] |

[a] Reference [54], 71 nm thick film, nanosecond pulse.

[b] Reference [49], 320 nm thick film, picosecond pulse.

[c] Reference [59], 21μm thick film on a BaF$_2$ substrate, picoseconds pulse.

[d] Reference [70], 60 nm thick film on a silica substrate.

[e] Reference [58], C$_{60}$ toluene solution, nanosecond pulse.

[f] Reference [57], C$_{60}$ benzene solution, picosecond pulse.

[g] Reference [71], C$_{60}$ toluene solution, nanosecond pulse.

[h] Reference [61], 10 μm amorphous film on a CaF$_2$ substrate, femtosecond pulse.

[i] Reference [72], DFT/LDA molecular response calculation.

[j] Reference [73], valence-effective-Hamiltonian SOS method.

[k] Reference [74], INDO/SCI and INDO/SDCI methods.

[l] Reference [56], Sum-Over-Orbitals method within local density approximation.

[m] Reference [75], Finite-Field method within local density approximation.

## 4. Conclusions

We have presented a many-body BSE-SOS method for the linear and nonlinear optical susceptibilities of solids. Excitonic and local field effects have been taken into account naturally. We directly calculate the linear OPA spectrum and nonlinear THG and DFWM spectra of solid $C_{60}$ fullerene instead of indirectly estimating the $\chi^{(3)}$ from the molecular $\gamma$ by local field correction. For OPA and THG, although a deviation in the position of the first characteristic peak exists between the theoretical and experimental results, the overall agreement is good in terms of line shape and position of the second characteristic peak. We have identified the type of resonances of the characteristic peaks by tracing the SOS process. Our theoretical insight provides a good reference to understand the discrepancy in identification of resonance for two characteristic peaks in two THG experiments [49,54,68]. For DFWM, a better agreement than THG is obtained between the theoretical and experimental results in terms of line shape and peak positions. We have shown that the BSE-SOS method is an advisable strategy to estimate the linear and nonlinear optical susceptibilities of solids. However, a comparison between the magnitudes of theoretical and experimental nonlinear susceptibilities is still difficult.

**Acknowledgements**

We appreciate the financial support from Natural Science Foundation of China Project 21303164 and State Key Laboratory of Structural Chemistry Project 20200017.

**References**:

[1]  C. Attaccalite, E. Cannuccia, and M. Grüning, Phys. Rev. B **95**, 125403 (2017).

[2]  E. K. Chang, E. L. Shirley, and Z. H. Levine, Phys. Rev. B **65**, 035205 (2001).

[3]  M. Grüning and C. Attaccalite, Phys. Rev. B **89**, 081102 (2014).

[4]  M. Grüning and C. Attaccalite, Phys. Rev. B **90**, 199901 (2014).

[5]  Y.-Z. Lan and X.-H. Bao, Phys. Rev. B **101**, 195437 (2020).

[6]  R. Leitsmann, W. G. Schmidt, P. H. Hahn, and F. Bechstedt, Phys. Rev. B **71**, 195209 (2005).

[7]  T. G. Pedersen, Phys. Rev. B **92**, 235432 (2015).

[8]  M. Rohlfing and S. G. Louie, Phys. Rev. B **62**, 4927 (2000).

[9]  J. L. Brédas, C. Adant, P. Tackx, A. Persoons, and B. M. Pierce, Chem. Rev. **94**, 243 (1994).

[10] B. Champagne and B. Kirtman, J. Chem. Phys. **125**, 024101 (2006).

[11] W. D. Cheng, D. S. Wu, H. Zhang, X. D. Li, Y. Z. Lan, D. G. Chen, and H. X. Wang, J. Chem. Phys. **119**, 13100 (2003).

[12] D. R. Kanis, M. A. Ratner, and T. J. Marks, Chem. Rev. **94**, 195 (1994).

[13] M. G. Kuzyk, Phys. Rev. A **72**, 053819 (2005).

[14] P. K. Nandi, N. Panja, and T. Kar, Chem. Phys. Lett. **444**, 366 (2007).

[15] B. M. Pierce, J. Chem. Phys. **91**, 791 (1989).

[16] D. P. Shelton and J. E. Rice, Chem. Rev. **94**, 3 (1994).

[17] W. Q. Tian, J. Comput. Chem. **33**, 466 (2012).

[18] F. Meyers, S. R. Marder, B. M. Pierce, and J. L. Brédas, J. Am. Chem. Soc. **116**, 10703 (1994).

[19] Y. Z. Lan, W. D. Cheng, D. S. Wu, X. D. Li, H. Zhang, and Y. J. Gong, Chem. Phys. Lett. **372**, 645 (2003).

[20] Y. Z. Lan, W. D. Cheng, D. S. Wu, J. Shen, S. P. Huang, H. Zhang, Y. J. Gong, and F. F. Li, J. Chem. Phys. **124**, 094302 (2006).

[21] X.-L. Zheng, L. Yang, B. Shang, M.-Q. Wang, Y. Niu, W.-Q. Li, and W. Q. Tian, Phys. Chem. Chem. Phys. **22**, 14225 (2020).

[22] C. Aversa and J. E. Sipe, Phys. Rev. B **52**, 14636 (1995).

[23] J. L. P. Hughes and J. E. Sipe, Phys. Rev. B **53**, 10751 (1996).

[24] D. J. Moss, E. Ghahramani, J. E. Sipe, and H. M. van Driel, Phys. Rev. B **41**, 1542 (1990).

[25] J. E. Sipe and E. Ghahramani, Phys. Rev. B **48**, 11705 (1993).

[26] J. E. Sipe and A. I. Shkrebtii, Phys. Rev. B **61**, 5337 (2000).

[27] K. S. Virk and J. E. Sipe, Phys. Rev. B **76**, 035213 (2007).

[28] Y.-Z. Lan, Comput. Cond. Mat. **8**, 22 (2016).

[29] S. N. Rashkeev, W. R. L. Lambrecht, and B. Segall, Phys. Rev. B **57**, 3905 (1998).

[30] S. Sharma, J. K. Dewhurst, and C. Ambrosch-Draxl, Phys. Rev. B **67**, 165332 (2003).

[31] R. Bursi, M. Lankhorst, and D. Feil, J. Comput. Chem. **16**, 545 (1995).

[32] J. P. Perdew, K. Burke, and M. Ernzerhof, Phys. Rev. Lett. **78**, 1396 (1997).

[33] Y. Mokrousov, G. Bihlmayer, and S. Blügel, Phys. Rev. B **72**, 045402 (2005).

[34] C. Attaccalite, arXiv e-prints arXiv:1609.09639 (2016).

[35] A. Taghizadeh, F. Hipolito, and T. G. Pedersen, Phys. Rev. B **96**, 195413 (2017).

[36] K. Dewhurst, S. Sharma, L. Nordström, F. Cricchio, F. Bultmark, O. Gräs, and H. Gross, (see

http://elk.sourceforge.net).

[37] R. W. Boyd, *Nonlinear Optics*, Second edition (Academic San Diego, CA., 2003), p. 135.

[38] P. Norman and K. Ruud, in edited by M. G. Papadopoulos, A. J. Sadlej, and J. Leszczynski, 1st ed. (Springer Netherlands, Springer Science+Business Media B.V., 2006), pp. 1–49.

[39] J. A. Delaire and K. Nakatani, Chem. Rev. **100**, 1817 (2000).

[40] C. Aversa, J. E. Sipe, M. Sheik-Bahae, and E. W. Van Stryland, Phys. Rev. B **50**, 18073 (1994).

[41] M. Rohlfing and J. Pollmann, Phys. Rev. B **63**, 125201 (2001).

[42] G. P. P. Giuseppe Grosso, *Solid State Physics, Second Edition*, 2nd ed. (Academic Press, 2013).

[43] A. Szabo and N. S. Ostlund, *Modern Quantum Chemistry: Introduction Advanced Electronic Structure Theory* (Dover Publications, 1996).

[44] P. A. Heiney, J. E. Fischer, A. R. McGhie, W. J. Romanow, A. M. Denenstein, J. P. McCauley Jr., A. B. Smith, and D. E. Cox, Phys. Rev. Lett. **66**, 2911 (1991).

[45] Computational results obtained using software programs from Accelrys. Geometry optimizations performed with the CASTEP program, and graphical displays generated with Materials Studio 4.0, Inc. San Diego, (2006).

[46] S. Jalali-Asadabadi, E. Ghasemikhah, T. Ouahrani, B. Nourozi, M. Bayat-Bayatani, S. Javanbakht, H. A. R. Aliabad, I. Ahmad, J. Nematollahi, and M. Yazdani-Kachoei, J. Electron. Mater. **45**, 339 (2016).

[47] E. Y. Kolyadina, L. A. Matveeva, P. L. Neluba, and E. F. Venger, Semiconductor Physics, Quantum Electronics & Optoelectronics **18**, 349 (2015).

[48] E. L. Shirley and S. G. Louie, Phys. Rev. Lett. **71**, 133 (1993).

[49] J. S. Meth, H. Vanherzeele, and Y. Wang, Chem. Phys. Lett. **197**, 26 (1992).

[50] S. Z. Karazhanov, P. Ravindran, A. Kjekshus, H. Fjellvåg, and B. G. Svensson, Phys. Rev. B **75**, 155104 (2007).

[51] E. Luppi, H. Hübener, and V. Véniard, Phys. Rev. B **82**, 235201 (2010).

[52] T. Rabenau, A. Simon, R. K. Kremer, and E. Sohmen, Z. Phys. B Cond. Mat. **90**, 69 (1993).

[53] J. H. Weaver, P. J. Benning, F. Stepniak, and D. M. Poirier, J. Phys. Chem. Solids **53**, 1707 (1992).

[54] F. Kajzar, C. Taliani, R. Danieli, S. Rossini, and R. Zamboni, Chem. Phys. Lett. **217**, 418 (1994).

[55] J. R. Lindle, R. G. S. Pong, F. J. Bartoli, and Z. H. Kafafi, Phys. Rev. B **48**, 9447 (1993).

[56] E. Westin and A. Rosén, Appl. Phys. A **60**, 49 (1995).

[57] W. J. Blau, H. J. Byrne, D. J. Cardin, T. J. Dennis, J. P. Hare, H. W. Kroto, R. Taylor, and D. R. M.

Walton, Phys. Rev. Lett. **67**, 1423 (1991).

[58] Q. Gong, Y. Sun, Z. Xia, Y. H. Zou, Z. Gu, X. Zhou, and D. Qiang, J. Appl. Phys. **71**, 3025 (1992).

[59] Z. H. Kafafi, J. R. Lindle, R. G. S. Pong, F. J. Bartoli, L. J. Lingg, and J. Milliken, Chem. Phys. Lett. **188**, 492 (1992).

[60] Y. Wang and L. T. Cheng, J. Phys. Chem. **96**, 1530 (1992).

[61] F. P. Strohkendl, T. J. Axenson, R. J. Larsen, L. R. Dalton, R. W. Hellwarth, and Z. H. Kafafi, J. Phys. Chem. B **101**, 8802 (1997).

[62] G. Banfi, D. Fortusini, M. Bellini, and P. Milani, Phys. Rev. B **56**, R10075 (1997).

[63] M. Fanti, G. Orlandi, and F. Zerbetto, J. Am. Chem. Soc. **117**, 6101 (1995).

[64] L. Geng and J. C. Wright, Chem. Phys. Lett. **249**, 105 (1996).

[65] F. Kajzar, C. Taliani, R. Zamboni, S. Rossini, and R. Danieli, Synth. Met. **77**, 257 (1996).

[66] P. Norman, Y. Luo, D. Jonsson, and H. Ågren, J. Chem. Phys. **106**, 8788 (1997).

[67] G. B. Talapatra, N. Manickam, M. Samoc, M. E. Orczyk, S. P. Karna, and P. N. Prasad, J. Phys. Chem. **96**, 5206 (1992).

[68] G. P. Zhang, X. Sun, and T. F. George, J. Phys. Chem. A **113**, 1175 (2009).

[69] S. Wang, N. Ye, W. Li, and D. Zhao, J. Am. Chem. Soc. **132**, 8779 (2010).

[70] H. Hoshi, N. Nakamura, Y. Maruyama, T. Nakagawa, S. Suzuki, H. Shiromaru, and Y. Achiba, Jpn. J. Appl. Phys. **30**, L1397 (1991).

[71] D. Neher, G. I. Stegeman, F. A. Tinker, and N. Peyghambarian, Opt. Lett. **17**, 1491 (1992).

[72] Y. Luo, P. Norman, P. Macak, and H. Ågren, Phys. Rev. B **61**, 3060 (2000).

[73] Z. Shuai and J. L. Brédas, Phys. Rev. B **48**, 11520 (1993).

[74] J. Li, J. Feng, and J. Sun, Chem. Phys. Lett. **203**, 560 (1993).

[75] N. Matsuzawa and D. A. Dixon, J. Phys. Chem. **96**, 6872 (1992).